# Performance Analysis of Clustering Protocol Using Fuzzy Logic for Wireless Sensor Network

**Vaibhav Godbole**

Departement of Information Technology, Fr. Conceicao Rodrigues College of Engineering,
Fr. Agnel Technical Education Complex, Bandra (W), Mumbai: 400050

---

| Article Info | ABSTRACT |
|---|---|

In order to gather information more efficiently, wireless sensor networks (WSNs) are partitioned into clusters. Most proposed clustering algorithms do not consider the location of the base station. This situation causes hot spot problems in multi-hop WSNs. In this paper, we analyze a fuzzy clustering algorithm (FCA) which aims to prolong the lifetime of WSNs. This algorithm adjusts the cluster-head radius considering the residual energy and distance to the base station parameters of the sensor nodes. This helps to decrease the intra-cluster work of the sensor nodes which are closer to the base station or have lower battery level. Fuzzy logic is utilized for handling the uncertainties in cluster-head radius estimation. We compare this algorithm with the low energy adaptive clustering hierarchy (LEACH) algorithm according to the parameters of first node dies half of the nodes alive and energy-efficiency metrics. Our simulation results show that the fuzzy clustering approach performs better than LEACH. Therefore, the FCA is a stable and energy-efficient clustering algorithm.

---

*Corresponding Author:*

Vaibhav Godbole
Departement Of Information Technology,
Fr. Conceicao Rodrigues College Of Engineering,
Fr. Agnel Technical Education Complex, Bandra (W), Mumbai: 400050.
Email: vai.godbole@gmail.com

---

## 1. INTRODUCTION

A wireless sensor network (WSN) [1] consists of spatially distributed autonomous sensors to cooperatively monitor physical or environmental conditions, such as temperature, sound, vibration, pressure, motion or pollutants. The development of WSNs was motivated by military applications such as battlefield surveillance. They are now used in many industrial and civilian application areas, including industrial process monitoring and control, machine health monitoring, environment and habitat monitoring, healthcare applications, home automation, and traffic control [2].

In addition to one or more sensors, each node in a sensor network is typically equipped with a radio transceiver or other wireless communications device, a small micro-controller, and an energy source, usually a battery. A sensor node might vary in size from that of a shoebox down to the size of a grain of dust, although functioning "motes" [3] of genuine microscopic dimensions have yet to be created. The cost of sensor nodes is similarly variable, ranging from hundreds of dollars to a few pennies, depending on the size of the sensor network and the complexity required of individual sensor nodes. Size and cost constraints on sensor nodes result in corresponding constraints on resources such as energy, memory, computational speed and bandwidth [4]. Figure 1 shows a typical WSN.

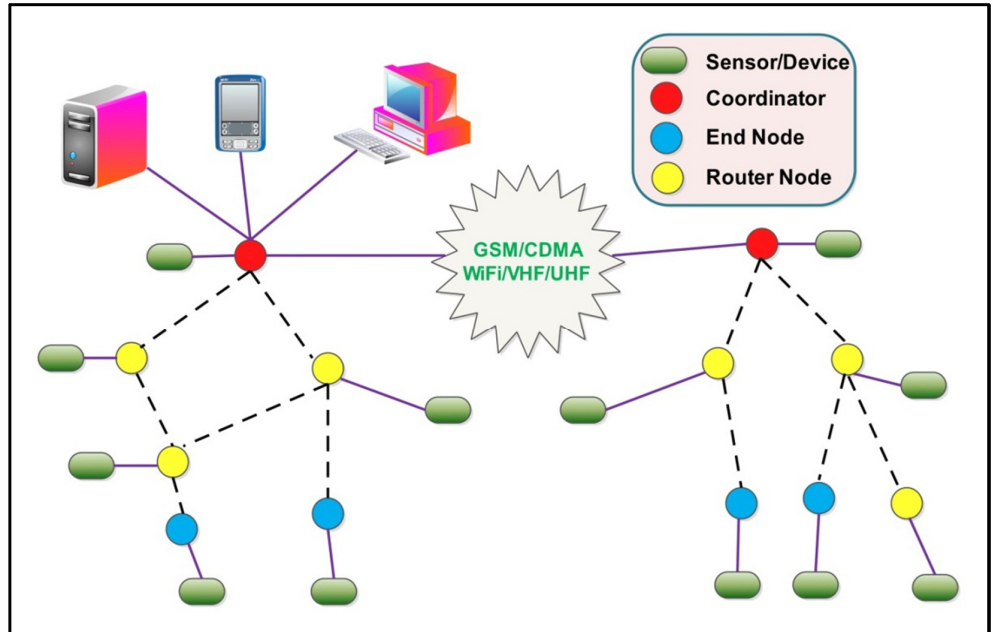

Figure 1. A typical WSN architecture [4].

Several WSN applications require only an aggregate value to be reported to the observer. In this case, sensors in different regions of the field can collaborate to aggregate their data and provide more *accurate* reports about their local regions. Data aggregation reduces the communication overhead in the network, leading to significant energy savings. In order to support data aggregation through efficient network organization, nodes can be partitioned into a number of small groups called *clusters*. Each cluster has a coordinator, referred to as a *cluster head*, and a number of *member* nodes [5].

Most clustering algorithms utilize two techniques which are selecting cluster-heads with more residual energy and rotating cluster-heads periodically to balance the energy consumption of the sensor nodes over the network . These clustering algorithms do not take the location of the base station into consideration. This lack of consideration causes hot spot problems in multi-hop WSNs [6]. The cluster-heads near the base station die earlier, because they will be in  heavier relay traffic than the cluster-heads which are relatively far from the base station.. In order to solve this problem and to balance energy consumption of cluster-heads, a periodically rotating cluster-head mechanism was proposed by Yu and Chang [5], namely low-energy adaptive clustering hierarchy (LEACH), which is a clustering algorithm that utilizes randomized rotation to balance energy consumption of cluster-heads over the network.

 LEACH uses a probabilistic approach to select the cluster-head schematically. Hoeever this method uses only local information to make decisions on the cluster-head so using only local information has its own limitations. Since each node probabilistically elects whether or not to become the cluster-head, there might be cases when two cluster-heads are selected in closed proximity of each other. In reality, considering only one factor such as energy is not suitable to select the cluster-head properly. This is because other conditions such as centrality of nodes corresponding to the entire cluster, gives an amount to the entire dissipation during transmission for all nodes too. The more central the node to a cluster the more is the energy efficiency for other nodes to transmit through that selected node. Periodical rotation is a vital property for clustering algorithms, it is not sufficient by itsel as incorporates delay in data packet transmission due to the movements of nodes being random most of the time [7].

In this paper, a fuzzy clustering approach [8] to the WSNs is analyzed to maximize its lifetime [9]. This approach is a distributed competitive algorithm. It selects the cluster-head via energy-based competition among the tentative cluster-heads which are selected using a probabilistic model. This approach mostly focuses on wisely assigning competition ranges to the tentative cluster-heads. In order to make wise decisions, it utilizes the residual energy and distance to the base station parameters of the sensor nodes. In addition to this, the clustering approach uses fuzzy logic to handle uncertainties in competition range estimation [8]. This allows the algorithm to assign greater competition ranges to the tentative cluster-heads which have higher residual energy levels, because they can serve a larger region [10].

## 2.    RELATED WORK - FUZZY CLUSTERING ALGOROTHMS

There are several clustering algorithms for WSNs in recent years [11]. Fuzzy logic is useful for making real time decisions without needing complete information about the environment [6]. On the other hand, conventional control mechanisms generally need accurate and complete information about the environment. Fuzzy logic can also be utilized for making a decision based on different environmental parameters by blending them according to predefined rules [12].

Some of the clustering algorithms [13] employ fuzzy logic to handle uncertainties in the WSNs. Basically, FCAs use fuzzy logic for blending different clustering parameters to select cluster-heads [8]. They assign chances to tentative cluster-heads according to the defuzzified output of fuzzy if-then rules. The tentative cluster-head becomes a cluster-head if it has the greatest chance in its vicinity. There are distributed and centralized fuzzy logic clustering approaches.

In the fuzzy clustering approach proposed by Rajashree et al. [14], the cluster-heads are selected at the base station. In every round, each sensor node forwards its clustering information to the base station. There are three fuzzy descriptors which are considered by the base station during cluster-head election which are node concentration, residual energy in each node and node centrality which are [14]:

1) Node concentration: Number of the nodes in the vicinity.
2) Residual energy: Remaining battery energy of each sensor node.
3) Node centrality: A parameter that indicates how central the node is to the cluster.

There are 27 fuzzy if-then rules which are defined at the base station [15]. The base station elects the cluster-heads according to these fuzzy rules. After the base station selects the cluster-head, it forwards the election results to entire network. This algorithm is a centralized clustering algorithm, because all clustering decisions are made at the base station. Gupta et al. [8] reported that a centralized clustering approach will produce more accurate cluster-heads, because the base station has all clustering information about the network and base stations are more powerful than ordinary nodes [16]. However, this centralized approach has some disadvantages [6, 13]:

1) The base station must collect all clustering information from the network. Repeating this process in every round brings a high overhead to sensor nodes. Thus, the battery levels of the sensor nodes may run low quickly.
2) In this approach the simulation is done for electing only one cluster-head per round. Therefore, this simulation is not a realistic one.

## 3.    LEACH CLUSTERING PROTOCOL

LEACH [17] is a distributed algorithm which makes local decisions to elect cluster-heads. If the cluster-heads are selected once and do not change throughout the network lifetime, then it is obvious that these static cluster-heads die earlier than the ordinary nodes. Therefore, LEACH includes randomized rotation of cluster-head locations to evenly distribute the energy dissipation over the network. LEACH also performs local data compression in cluster heads to decrease the amount of data that is forwarded to the base station.

LEACH, cluster-head selection is done periodically to enable randomized rotation of cluster-heads. Every round consists of two phases, namely set-up phase and steady-state phase. In the set-up phase, cluster-heads are selected and clusters are formed. In the steady-state phase, data transfers to the base station are performed through the clustered network [18]. A particular sensor node decides whether it is going to become a cluster-head or not by generating a random number between 0 and 1. If this number is less than the predefined threshold $T$ (n), then the sensor node becomes a cluster-head. G represents the set of sensor nodes that have not been cluster-heads in the last    where $P$ is the desired percentage of cluster-heads and $r$ represents the current round number. If the sensor node $n$ belongs to $G$ using these parameters, $T$ (n) is formulated as follows [19]:

$$T(n) = \frac{P}{1 - P * (r \mod \frac{1}{P})}$$

(1)

If the sensor node n does not belong to G, then the $T$ (n) is set to 0. Thus, n cannot become a cluster-head. At round 0, the probability of becoming a cluster-head for each node is equal to P. However, this situation changes in the following rounds. The cluster-heads of round 0 cannot become cluster-heads during the following *1 / P* rounds. This restriction prevents a particular node to become a cluster-head frequently. However, this restriction brings a drawback: it causes rapid decrease in the number of cluster-heads. To handle this drawback, as r increases, the chance of the remaining sensor nodes to be a cluster-head is also increased by adjusting the threshold $T$ (n) for the remaining sensor nodes. This critical balance is a significant property of LEACH [19].

After cluster-heads are selected for a particular round, each cluster-head broadcasts an advertisement message to the remaining sensor nodes. As each non-cluster-head node receives these advertisement messages, they decide the cluster to which they belong. Each non-cluster-head joins to the cluster from which it has received the largest signal strength. In order to join to the selected cluster, it transmits a *JoinClusterHeadMessage* to that cluster [20]. Once all the cluster-heads are selected and the clusters are formed, data transmission continues up to the next round. The simulations in Zhang and Wei [17] show that LEACH reduces communication energy as much as eight times as compared to direct transmission. In other words, the first node death in LEACH occurs eight times later than the first node death in direct transmission.

## 4.    APPLICATION OF FCA TO WSN

Fuzzy clustering algorithm is explained below (Algorithm 1). In every clustering round, each sensor node generates a random number between 0 and 1 [21]. If the random number for a particular node is smaller than the predefined threshold $T$ which is the percentage of the desired tentative cluster-heads, then that sensor node becomes a tentative cluster-head. In FCA the competition radius of each tentative cluster-head changes dynamically. FCA uses residual energy [22] parameter with distance to the base station metric of the sensor node to calculate competition radius. It is logical to decrease the service area of a cluster-head while its residual energy is decreasing. If the competition radius does not change as the residual energy decreases, the sensor node runs out of battery rapidly. This approach takes this situation into consideration and decreases the competition radius of each tentative cluster-head as the sensor node battery level decreases. Radius computation is accomplished by using predefined fuzzy if-then mapping rules [23] to handle the uncertainty. These fuzzy if-then mapping rules are given in Table 1.  The Mamdani method used by Kim et al. [24] is used as fuzzy inference technique, because it is the most frequently used fuzzy inference technique.

```
1:   T ← probability to become a tentative cluster-head
2:   nodeState ← CLUSTERMEMBER
3:   clusterMembers ← empty
4:   myClusterHead ← this
5:   beTentativeHead ← TRUE
6:   μ ← rand(0,1)
7:   IF μ < T THEN
8:      Calculate R_comp using fuzzy if-then mapping rules
9:      Advertise CandidateClusterHeadMessage (ID, R_comp, residualEnergy)
10:     On receiving CandidateClusterHeadMessage from node N
11:     if this.residualEnergy < N.residualEnergy then
12:       beTentativeHead ← False
13:       Advertise QuitElectionMessage(ID)
14:     end if
15:  end if
16:  if beT entativeHead = TRUE then
17:     Advertise ClusterHeadMessage(ID)
18:     ← CLUSTERHEAD
19:     On receiving JoinClusterHeadMessage(ID) from node N
20:     add N to the clusterMembers list
21:     EXIT
22:  else
23:     On receiving all ClusterHeadMessages
24:     myClusterHead ← the closest cluster-head
25:     Send JoinClusterHeadMessage(ID)
26:     EXIT
27:  end if
```

Algorithm 1. The proposed FCA for WSN.

Table 1. Fuzzy if-then mapping rules for competition radius calculation in FCA.

| Rule No. | Distance to Base | Residual Energy | Competotion Radius |
|----------|-----------------|-----------------|--------------------|
| 1 | Close | Low | Very Small |
| 2 | Close | Medium | Small |
| 3 | Close | Medium | Small |
| 4 | Medium | Low | Small |
| 5 | Medium | Medium | Medium |
| 6 | Medium | Medium | Large |
| 7 | Far | Low | Large |
| 8 | Far | Medium | Large |
| 9 | Far | High | Very Large |

In this approach for cluster-head competition radius calculation, we use two fuzzy input variables. The first one is the distance to the base station of a particular tentative cluster-head. The fuzzy set that describes the distance to base the station input variable is depicted in Figure 2. The linguistic variables for this fuzzy set are *close*, *medium* and *far*. We choose a trapezoidal membership function for *close* and *far* [25]. On the other hand, the membership function of medium is a triangular membership function.

The second fuzzy input variable is residual energy of the tentative cluster-head. The fuzzy set that describes residual energy input variable is illustrated in Figure 3. *Low*, *medium* and *high* are the linguistic variables of this fuzzy set. *Low* and *high* linguistic variables have a trapezoidal membership function while medium has a triangular membership function [26].

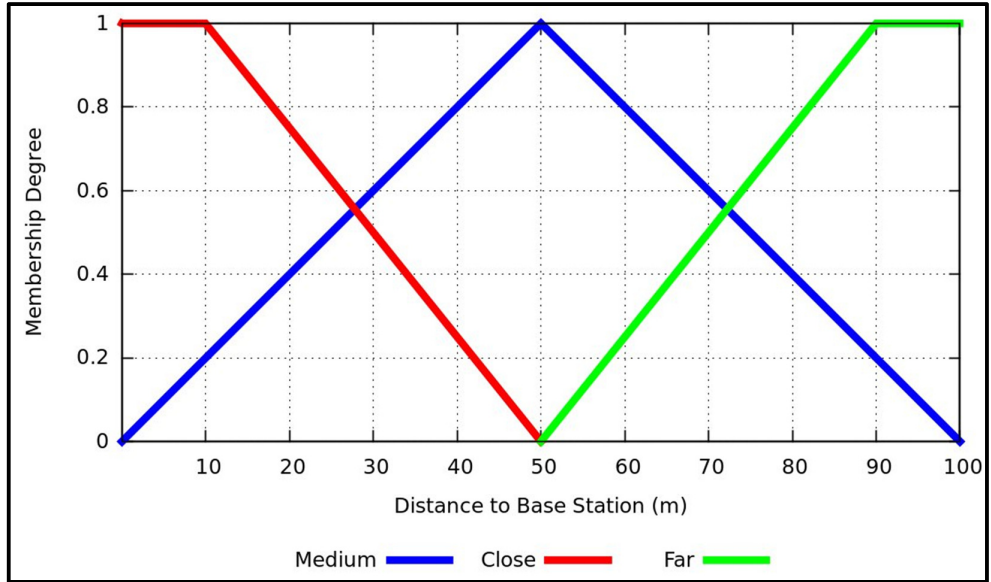

Figure 2. Fuzzy set for fuzzy input variable *Distance to Base.*

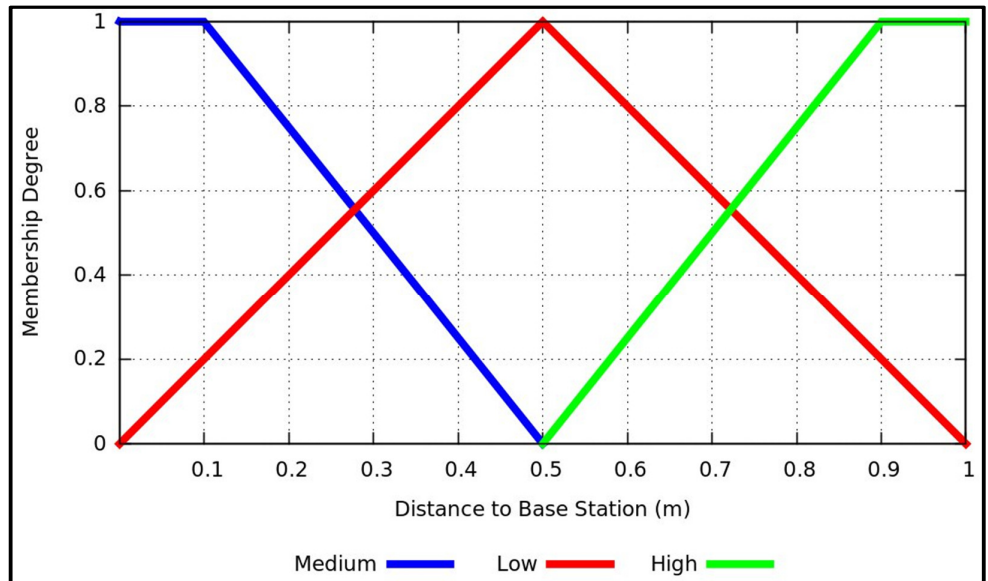

Figure 3. Fuzzy set for fuzzy input variable *Residual Energy.*

The only fuzzy output variable is the competition radius of the tentative cluster-head. Fuzzy set for competition radius fuzzy output variable is demonstrated in Figure 3. There are linguistic variables which are *very small, small, medium, large* and *very large. Very small* and *very large* have a trapezoidal membership function. The remaining linguistic variables are represented by using triangular membership functions.

If a particular tentative cluster-head's battery is full and it is located at the maximum distance to the base station, then it has the maximum competition radius [27]. On the contrary, if a particular cluster-head's battery is nearing empty and is the closest node to the base station, then it has the minimum competition radius. The remaining intermediate possibilities fall between these two extreme cases.

The maximum competition radius is a static parameter for a particular WSN. The base station broadcasts the value of this parameter to the entire network [28]. Thus, all the sensor nodes know the maximum competition radius, in advance. Each of the sensor nodes can calculate their relative competition radius according to the value of this parameter. The maximum distance to the base station is also a static parameter, because we assume that the sensor nodes are stationary. Each sensor node can determine their relative position to the base station considering the maximum distance to the base station in the WSN [19].

## 5. RESULTS AND DISCUSSION

We compare proposed FCA with LEACH using WiseNet Simulator, which is an open source Java based tool used to simulate the sensor network topology with secured protocols. WiseNet has the following features [29]:

- Graphical interface for reading the results of important parameters (eg. power consumption, reliability, coverage of the network.)
- Modeling types of attacks, programming of malicious sensors behavior and measuring the impacts on the parameters mentioned above.

In each round of the scenario, clusters-heads are selected and clusters are formed. Next, each ordinary node forwards certain bits of data to its cluster-head. Each cluster-head aggregates the received data and forwards it to the base station with a particular routing protocol or directly transmits the aggregated data to the base station. LEACH cluster-heads transmit their data packets to the base station directly. The area of deployed wireless sensor network is same for all scenarios and is 200 m x200 m. In each round, each ordinary sensor node transmits 4000 bits of data to its cluster-head. The cluster-head which receives the data from its cluster members aggregates the received data with a certain aggregation ratio. This aggregation ratio is set to 10% in our simulations.

In order to produce more reliable results, every scenario is simulated for 50 times, and the average of the results is taken. For each of the scenarios, we provide a summary result table which represents the values of first node dead (FND) and half node alive (HNA) [30] metrics for each of the algorithms simulated. After that, we provide a summary chart which illustrates the values of FND and HNA metrics visually. We also generate charts for the distribution of the number of live sensor nodes and the distribution of the number of clusters per each round. By using these simulation results, we comment on the performance of the simulated algorithms.

### 5.1. Scenario 1

In this scenario, the base station is located at the center of the WSN [3]. Each cluster-head forwards the aggregated data to the base station directly without using a relay node. The detailed configuration of this scenario is depicted in Table 2. The simulation of this scenario yielded the results shown in Table 3, which shows the rounds in which the FND and HNA for each simulated algorithm.

Table 2. Configuration parameters of Scenario 1.

| Parameter | Value |
|---|---|
| Network Size | 200 m  X 200 m |
| Base Station Location | (100,100) m |
| Num. Of Sensor Nodes | 75 |
| Initial Energy | 1 J |
| Data Packet Size | 3000 B |
| Aggregation Ratio | 10 % |

Table 3. Scenario 1: Values of FND and HNA metrics for each algorithm.

| Algorithm | FND | HNA |
|---|---|---|
| LEACH | 280 | 610 |
| FCA | 420 | 810 |

As seen in Table 3, the FCA performs better than LEACH. This algorithm is more efficient than LEACH by about 36.4%. LEACH's performance is poor because it does not consider the residual energy level of the sensor nodes during clustering. It uses a pure probabilistic model for clustering, but this model itself is not sufficient for providing the best solution.

Figure 5 depicts the distribution of the number of live sensor nodes with respect to the number of rounds for each algorithm. This figure clearly depicts the number of non-functional (dead) sensor nodes for fuzzy clustering approach begins after LEACH algorithm.

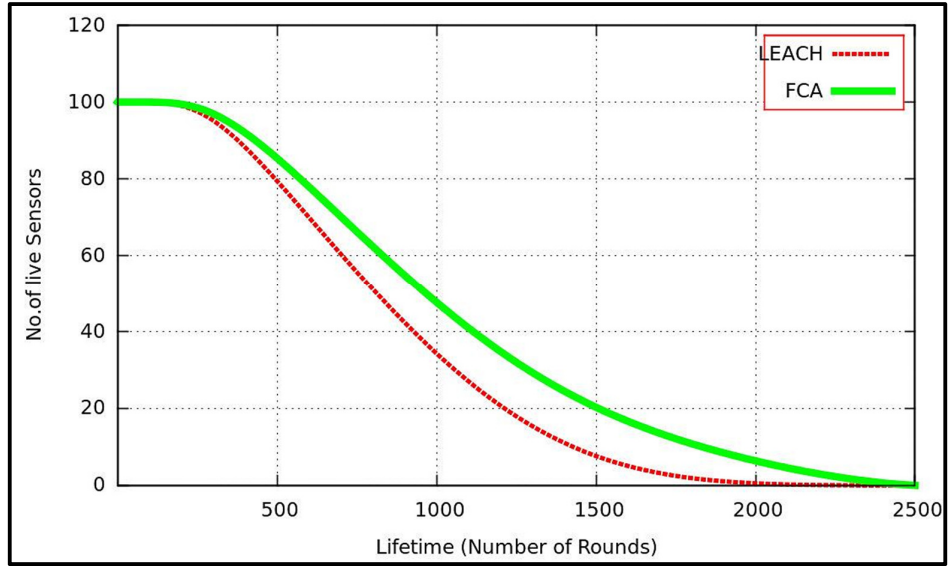

Figure 5. Scenario 1. Distribution of alive sensor nodes according to the number of rounds for each algorithm.

### 5.2. Scenario 2

In this scenario, the density of the deployed sensor nodes is doubled with respect to Scenario 1. We aim to test the behaviours of the clustering algorithms in different sensor network topologies which have different number of deployed sensor nodes which is similar to used by Gupta and Sampali [8]. In other words, we try to find out how clustering algorithms perform in relatively dense and sparse sensor network deployments. The detailed configuration of this scenario is illustrated in Table 4. The simulation of this scenario yielded the results shown in Table 5 which shows the rounds in which the FND and HNA for each simulated algorithm.

Table 4. Configuration parameters of Scenario 2.

| Parameter | Value |
|---|---|
| Network Size | 200m X 200m |
| Base Station Location | (100,100) m |
| No. Of Sensor Nodes | 150 |
| Initial Energy | 1 J |
| Data Packet Size | 3000 B |
| Aggregation Ratio | 10 % |

Table 5. Scenario 2: Values of FND and HNA metrics for each algorithm.

| Algorithm | FND | HNA |
|-----------|-----|-----|
| LEACH | 400 | 790 |
| FCA | 790 | 980 |

As seen in Table 5 the HNA performance of LEACH increased significantly in this scenario. LEACH sensor nodes start to die earlier than the sensor nodes FCA. FCA is more efficient than LEACH about 84.7 %

Figure 6 shows the distribution of live sensor nodes according to the number of rounds for each simulated algorithm. The number of sensor nodes of FCA is significantly greater than the other algorithms when the number of live sensor nodes is 100. This situation implies that FCA keeps WSN stable for a longer time than LEACH.

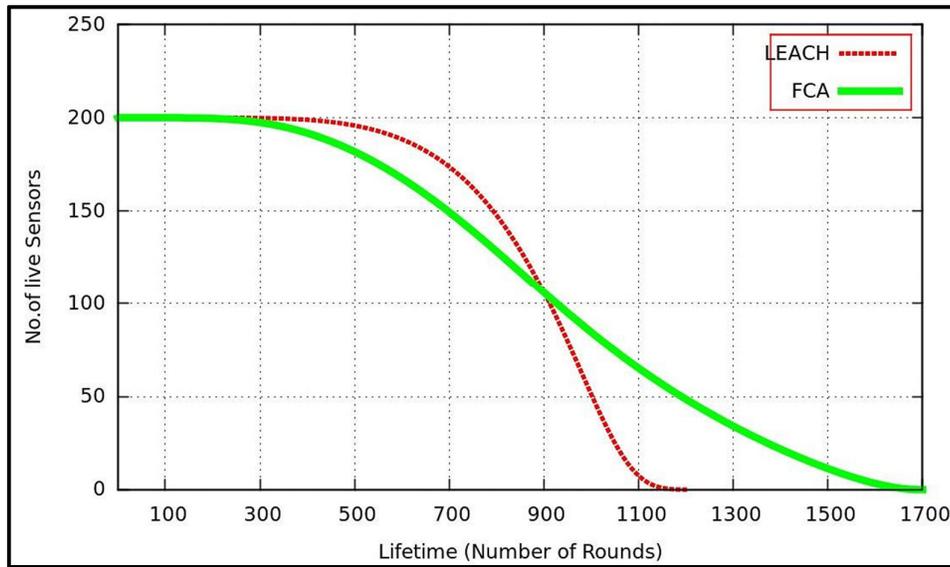

Figure 6. Scenario 2: Distribution of live sensor nodes according to the number of rounds for each algorithm.

## 6.    CONCLUSION AND FUTURE WORK

As a result of these experiments, we conclude that FCA is a stable and energy-efficient clustering algorithm for WSNs. This algorithm is designed for the WSNs that have stationary sensor nodes. As a future work, the fuzzy clustering approach can be extended for handling mobile sensor nodes.

Residual energy, distance to the base station and competition radius fuzzy sets can be adjusted in order to find optimal cluster-head radius values. In addition to this, the optimal maximum competition radius values for each scenario can be estimated by applying extensive tests. Some additional parameters such as node degree, density and local distance may also be employed to improve the performance of fuzzy clustering approach.

## ACKNOWLEDGEMENTS

I would like to thank the management of Fr. Conceicao Rodrigues College of Engineering, Bandra (W), Mumbai: 400050, for allowing me to use the infrastructure to perform the experimentation.

**BIBLIOGRAPHY OF AUTHORS**

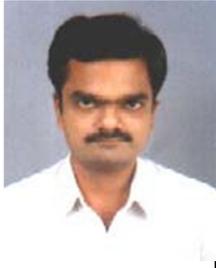

Vaibhav Godbole. Education: M.E. Electronics & Telecommunication (Mumbai University), B.E. Electrical Engineering (Mumbai University), DIE, DME. He is currently working as a Assistant Professor at Fr. Conceicao Rodigues College Of Engineering, Mumbai. His two research papers are in the press of Inderscience Publication Journal. He is also a reviewer for IET Journal of Networks. His areas of interests are mobile & wireless communications, evolutionary algorithms, genetic algorithms, new algorithms for mobile ad-hoc networks.